\documentclass[12pt]{article}

\usepackage[T1]{fontenc}
\usepackage[active]{srcltx}
\usepackage{lmodern}
\usepackage{amsmath,amstext,amssymb,graphicx,graphics,color}
\usepackage{theorem}
\usepackage{cite}               
\usepackage{hyperref}  
\usepackage{dsfont}             
\usepackage{fancybox}           
\usepackage{slashbox}           
\usepackage{mathdots}           
\usepackage[only,llbracket,rrbracket]{stmaryrd}

\theorembodyfont{\normalfont\slshape} 

\theoremstyle{break} 

{{\par\noindent \bf Proof. \nobreak}}%
{\nobreak \removelastskip \nobreak \hfill $\Box$ \medbreak}
{{\par\noindent \bf Proof \nobreak}}%
{\nobreak \removelastskip \nobreak \hfill $\Box$ \medbreak}
{{\par\noindent \bf Proof lemma. \nobreak}}%
{\nobreak \removelastskip \nobreak \bf End proof lemma. \medbreak}

\usepackage{fancyhdr}
 \fancypagestyle{plain}{
  \fancyhead{}
  
}

\def\paragraph#1{{\bf #1\ }}

\usepackage[utf8]{inputenc}

\title{Trail formation based on directed pheromone deposition}
\author{Emmanuel Boissard$^{(1,2)}$ \and Pierre Degond$^{(1,2)}$ \and Sebastien Motsch$^{(3)}$ }

\begin{document}

\maketitle
\vspace{0.5 cm}

\begin{center}
  1-Université de Toulouse; UPS, INSA, UT1, UTM ;\\ 
  Institut de Mathématiques de Toulouse ; \\
  F-31062 Toulouse, France. \\
  2-CNRS; Institut de Mathématiques de Toulouse UMR 5219 ;\\ 
  F-31062 Toulouse, France.\\
  email: emmanuel.boissard@math.univ-toulouse.fr ; pierre.degond@math.univ-toulouse.fr
\end{center}

\begin{center}
  3-Department of Mathematics\\ 
  University of Maryland\\
  College Park, MD 20742-4015\\
  email: smotsch@cscamm.umd.edu
\end{center}


\begin{abstract}
  We propose an Individual-Based Model of ant-trail formation. The ants are modeled as self-propelled particles which deposit directed pheromones and interact with them through alignment interaction. The directed pheromones intend to model pieces of trails, while the alignment interaction translates the tendency for an ant to follow a trail when it meets it. Thanks to adequate quantitative descriptors of the trail patterns, the existence of a phase transition as the ant-pheromone interaction frequency is increased can be evidenced. Finally, we propose both kinetic and fluid descriptions of this model and analyze the capabilities of the fluid model to develop trail patterns. We observe that the development of patterns by fluid models require extra trail amplification mechanisms that are not needed at the Individual-Based Model level.
\end{abstract}

\medskip
\noindent {\bf Acknowledgements:} The authors wish to thank J. Gautrais, C. Jost and G. Theraulaz for fruitful discussions. P. Degond wishes to acknowledge the hospitality of the Mathematics Department of Tsinghua University where this research was completed. The work of S. Motsch is partially supported by NSF grants DMS07-07949, DMS10-08397 and FRG07-57227.

\medskip
\noindent
{\bf Key words: } Self-propelled particles, pheromone deposition, directed pheromones, alignment interaction, Individual-Based Model, trail detection, pattern formation, kinetic models, fluid models

\medskip
\noindent
{\bf AMS Subject classification: } 35Q80, 35L60, 82C22, 82C31, 82C70, 82C80, 92D50.
\vskip 0.4cm

\setcounter{equation}{0}
\section{Introduction}
\label{sec:intro}

One of the many features displayed by self-organized collective motion of animals or individuals is the formation of trails. For instance, ant displacements are characterized by their organization into lanes consisting of a large number of individuals, for the purpose of exploring the environment or exploiting its resources. Another example involving species with higher cognitive capacities is the formation of mountain trails by hikers or herds of animals. In both cases, the main feature is that the interaction between the individuals is not direct, but instead, is mediated by a chemical substance or by the environment. Indeed, ants lay down pheromones as chemical markers. These pheromones are sensed by other individuals which use them to adjust their path. In the case of mountain trails, the modification of the soil by walkers facilitates the passage of the next group of individuals and attracts them. This phenomenon is well-known to biologists under the name of stigmergy, a concept first forged by Pierre-Paul Grass\'e \cite{grasse_termitologia:_1986} to describe the coordination of social insects in nest building.

The formation of trails by ants has been widely studied in the biological literature \cite{beckers_trail_1992, couzin_self-organized_2003, detrain_influence_2001, edelstein-keshet_simple_1994, edelstein-keshet_trail_1995, watmough_modelling_1995}.  One general observation is the fact that trail formation is a self-organized phenomenon and expresses the emergence of a large-scale order stemming from simple rules at the individual level. Indeed, ant colonies in the numbers of thousands of individuals or more arrange into lines without resorting to long-range signaling or hierarchical organization. Another striking feature is the variability of the trail patterns, which may range from densely woven networks to a few large trails. This flexibility may result from the ability of the individuals to adapt their activity to variable external conditions such as food availability, temperature, terrain conditions, the presence of predators, etc. Trail plasticity derives from internal and external factors: for example it may vary according to the species of ant under consideration or depends on the properties of the soil.  Our goal is to provide a model that accounts for these two general facts: spontaneous formation of trails, and variability of the trail pattern.

At the mathematical level, several types of ant displacement and pheromone deposition models have been introduced. A first series of works deal with ant displacement on a pre-existing pheromone trail and focus on the role of the antennas in the trail sensing mechanism \cite{calenbuhr_model_1992,calenbuhr_model_1992-1, couzin_self-organized_2003}. Spatially one-dimensional models do not specifically address the question of trail build-up either, since motion occurs on a one-dimensional predefined trail. One-dimensional cellular automata models have been used to determine the fundamental diagram of pheromone-regulated traffic and to study the spontaneous break-up of bi-directional traffic in one preferred direction \cite{john_collective_2004,nishinari_modelling_2006}.

The decision-making  mechanisms which lead to the selection of a particular branch when several routes are available have been modeled by considering Ordinary Differential Equations (ODE's) for the global ant and pheromone densities on each trail \cite{beckers_collective_1990,deneubourg_self-organizing_1990,goss_self-organized_1989,peters_analytical_2006}. These models do not account for the spontaneous formation of the trails. In \cite{edelstein-keshet_simple_1994}, the spatial distribution of trails is ignored in a similar way. However, it introduces the concept of a space-averaged statistical distribution of trails, which reveals to be very effective. In the present work, we have borrowed from \cite{edelstein-keshet_simple_1994} the idea of considering trails as particles in the same fashion as ants, and of dealing with them through the definition of a trail distribution function. However, by contrast to \cite{edelstein-keshet_simple_1994}, we keep track of both the spatial and directional distribution of these trails.

In general, two dimensional models consider that ant motion occurs on a fixed lattice. Two classes of ant models have been considered: Cellular Automata models \cite{ermentrout_cellular_1993, edelstein-keshet_trail_1995, watmough_modelling_1995}, and Monte-Carlo models \cite{beckers_collective_1990,deneubourg_self-organizing_1990,rauch_pattern_1995,schweitzer_active_1997,tao_flexible_2004}. In the first class of models, no site can be occupied by more than one ant, while in the second class, ants are modeled as particles subject to a biased random walk on the lattice. In \cite{vincent_effect_2004}, the authors introduce some mean-field approximation of the previous models: a time-continuous Master equation formalism is used to determine the evolution of the ant density on each edge. In all these models, the jump probabilities are modified by the presence of pheromones. The pheromones can be located on the nodes  \cite{schweitzer_active_1997,tao_flexible_2004,rauch_pattern_1995}, but the trail reinforcement mechanism seems more efficient if they are located on the edges \cite{ermentrout_cellular_1993, beckers_collective_1990,deneubourg_self-organizing_1990, edelstein-keshet_trail_1995, watmough_modelling_1995}. To enhance the trail formation mechanisms, some authors  \cite{schweitzer_active_1997,tao_flexible_2004} introduce two sorts of pheromones, an exploration pheromone which is deposited during foraging and a recruit pheromone which is laid down by ants who have found food and try to recruit congeners to exploit it. In  \cite{rauch_pattern_1995}, it is demonstrated that trail formation is enhanced by introducing some saturation of the ant sensitivity to pheromones at high pheromone concentrations. Inspired by the observation that pheromone deposition on edges seems to be more efficient in producing self-organized trails, we suppose that laid down pheromones give rise to trails (i.e. directed quantities) rather than substance concentration (i.e. scalar quantities). 

All these previously cited two-dimensional models assume a pre-existing lattice structure. One questions which is seldom addressed is whether this pre-existing lattice may influence the formation of the trails. For instance, it is well-known that lattice Boltzmann models with too few velocities have incorrect behavior. One may wonder if similar effects could be encountered with spatially discrete ant trail formation models. For this reason, in the present work, we will depart from a lattice-like spatial organization and treat the motion of the ants in the two-dimensional continuous space. 

In this work, we propose a time and space continuous Individual-Based Model for self-organized trail formation. In this model, self-propelled particles interact by laying down pheromone trails that indicate both the position and direction of the trails. Ants adapt their course by following trails deposited by others, therefore reinforcing existing trails while evaporation of pheromones allows weaker trails to disappear. The ant dynamics is time-discrete and is a succession of free flights and velocity jumps occuring at time intervals $\Delta t$. Velocity jumps occur with an exponential probability. Two kinds of velocity jumps are considered: purely random jumps which translate the ability of ants to explore a new environment, and trail-recruitment jumps. In order to perform the latter, ants look for trails in a disk around themselves, pick up one of these trails with uniform probability and adopt the direction of the chosen trail.

This model bears analogies with chemotaxis models. Chemotaxis is the name given to remote attraction interaction through chemical signaling in colonies of bacteria. Mathematical modeling of chemotaxis has been largely studied. Macroscopic models were first introduced by Keller and Segel in the form of a set of parabolic equations \cite{keller_model_1971}. These equations can be obtained as macroscopic limits of kinetic models \cite{erban_individual_2004, filbet_derivation_2005,hillen_diffusion_2000,othmer_diffusion_2002,othmer_aggregation_1997}. Kinetic models describe the evolution of the population density in position-velocity space.  In \cite{stevens_derivation_2000} a direct derivation of the Keller-Segel model from a stochastic many-particle model is given. The common feature of most chemotaxis models is the appearance of blow-up, which corresponds to the fast aggregation of individuals at a specific point in space (see e.g.  \cite{blanchet_two-dimensional_2006,calvez_volume_2006}). By contrast, in the present paper, the dynamics gives rise to the spontaneous organization of lane-like spatial patterns, much alike to the observed behavior of ants. The reason for this different morphogenetic behavior is the directed nature of the mediator of the interaction.

The model also bears analogies with the kinetic model of cell migration developed in \cite{painter_modelling_2009}. In this model, cells move in a medium consisting of interwoven extra-cellular fibers in the direction of one, randomly chosen fiber direction. As they move, cells specifically destroy the extracellular fibers which are transverse to their motion. The induced trail reinforcement mechanism produces a network. There are two differences with the trail formation mechanism that we present here. The first one is that the dynamics starts from a prescribed set of motion directions and gradually reduces this set, while our algorithms builds up the set of available directions gradually and new directions are created through random velocity jumps. The second difference is the role of trail evaporation in our algorithm, which has no equivalent in the cell motion model. Indeed, trail evaporation is a major ingredient for network plasticity.

In the last part of the present work, we derive a kinetic formulation of the proposed ant trail formation model in the spirit of \cite{painter_modelling_2009}. Then, the fluid limit of this kinetic model is considered. We show that the resulting fluid model can exhibit trail formation only if some concentration mechanism is involved, while numerical simulations indicate that trail formation may occur without such a mechanism. Therefore, the appearance of trails is enhanced when the model provides more information about the ant velocity distribution function.

The outline of this paper is as follows. In section \ref{sec:model}, we provide the model description.  Section \ref{sec:simulation} is devoted to the analysis of the numerical simulations. We establish a methodology for the detection of trail patterns from a simulation outcome, and analyze the dependency of the observed features on the model parameters. In section \ref{sec:kinetic}, we formally establish a set of kinetic equations that describes the dynamics and we investigate their fluid limit. A conclusion in section \ref{sec:conclu} draws some perspectives of this work.

\setcounter{equation}{0}
\section[An Individual-Based-Model of ant behavior]{An Individual-Based-Model of ant
  behavior \\based on directed pheromone deposition}
\label{sec:model}

\begin{figure}
  \centering
  \includegraphics[scale=.5]{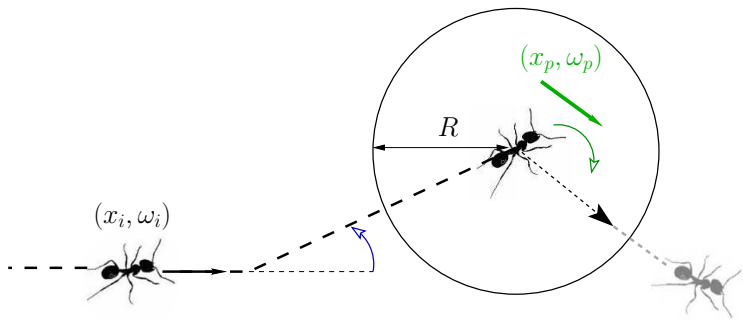}
  \caption{Ants follow a random walk process. Each ant is moving on a straight line until it undergoes a random velocity jump (left) or a trail recruitment jump (right). In this picture, a trail pheromone is located in the disk centered at the ant location when the jump occurs, and of given radius $R$.}
  \label{fig:ant_saut}
\end{figure}

We consider $N$ ``ants'' in a flat (2-dimensional) domain: each ant is described by its position $x_i\in \mathbb{R}^2$ and the direction of its motion $\omega_i$. The vector $\omega_i$ is supposed to be of unit-length, i.e. $\omega_i \in {\mathbb S}^1$, where ${\mathbb S}^1$ denotes the unit circle. We also consider pieces of trails described by pairs $(y_p,\,\omega_p)$ where $y_p\in \mathbb{R}^2$ is the trail piece position and $\omega_p$ is a unit vector describing the trail direction (see figure \ref{fig:ant_saut}). In the case of ants, the marking of the trail is realized by a chemical marker, namely a trail pheromone. We assume that the ants can distinguishably perceive the direction of the trail of this chemical marker and that they are able to follow, not the line of steepest gradient, like in chemotaxis, but the direction of this trail. Note that in the case of walkers or sheep in an outdoor terrain, the marking of the trail is realized by the modification of the terrain consecutive to the passage of the walkers, such as flattened grass. For wild white bears, this modification is realized by the trail left in the snow by the animals. In the sequel, we concentrate on the modeling of ant trail formation, and we will indistinguishably refer to these pieces of trails as 'trails' or 'trail pheromones', or simply, 'pheromones'. The set of pheromones varies with time, since new pheromones are created by the deposition process and pheromones disappear after some time in order to model the evaporation process. We will denote by ${\mathcal P}(t)$ the set of pheromones at time $t$.

The simulated ants follow a random walk process. During free flights, Ant $i$ moves in direction $\omega_i$ at a constant speed $c$, i.e. is subject to the differential equation:
\begin{equation}
  \dot x_i = c \omega_i, \quad \dot{\omega}_i = 0, \label{eq:free_flight}
\end{equation}
where the dots stand for time derivatives. This free motion is randomly interrupted by velocity jumps. When Ant $i$ undergoes a velocity jump at time $t$, its velocity direction before the jump $\omega_i(t-0)$ is suddenly changed into a different one $\omega_i (t+0)$. The jump times are drawn according to Poisson distributions. In practice, a time discrete algorithm is used, with time steps $\Delta t$. With such a discretization, a Poisson process of frequency $\lambda$ is represented by an event occuring with probability $1 - e^{- \lambda \Delta t}$ over this time step. There are two kinds of jumps: random velocity jumps and trail recruitment jumps. 

\medskip
\noindent
{\it Random velocity jumps.}  In this case, $\omega_i (t+0)$ differs from $\omega_i (t-0)$ by a random angle $\varepsilon$, i.e. $ \widehat{(\omega_i(t-0), \omega_i(t+0))} = \varepsilon$, where $\varepsilon$ is drawn out of a Gaussian distribution $p(\varepsilon)$ with zero mean and variance $\sigma^2$, periodized over $[0,2\pi]$, i.e. 
\begin{displaymath}
  p(\varepsilon) \, d \varepsilon = \sum_{n \in {\mathbb Z}} \frac{1}{(2 \pi \sigma^2)^{1/2}} \exp \left(-\frac{(\varepsilon + 2 n \pi)^2}{2 \sigma^2} \right) \, d \varepsilon, \quad \varepsilon \in [0,2\pi].
\end{displaymath}
The frequency of the Poisson process is constant in time and denoted by $\lambda_r$. 

\medskip
\noindent {\it Trail recruitment jump.} In this case, $\omega_i(t+0)$, is picked up with uniform probability among the directions $\omega_p$ of the trail pheromones located in the ball $B_R(x_i(t))$ of radius $R$ centered at $x_i(t)$ (see figure \ref{fig:ant_saut}). $B_R(x_i(t))$ is the ant detection region and $R$ its detection radius. More precisely, defining the set
\begin{displaymath}
  S_i(t) = \{ p \in {\mathcal P}(t) \, , \, |x_p - x_i(t)| \leq R \},
\end{displaymath}
Ant $i$ chooses an index $p$ in $S_i(t)$ with uniform probability and sets 
\begin{equation}
  \label{ew:polar_interaction}
  \omega_i(t+0)=\omega_p.
\end{equation}
A variant of this mechanism involves a nematic interaction (i.e. the deposited trails have no specific orientation).   In this case, the new direction is defined by 
\begin{equation}
  \label{ew:nematic_interaction}
  \omega_i(t+0)= \pm \omega_p, \quad \mbox{ such that } \quad \widehat{(\omega_i(t+0), \omega_p)} \quad \mbox{ is acute}. 
\end{equation}
The nematic interaction makes more biological sense, since it seems difficult to envision a mechanism which would allow the ants to detect the orientation of a given trail. The use of a uniform probability to select the interacting pheromone can be questioned. For instance, the choice of the trail pheromone $p$ could be dependent on the angle $\widehat{(\omega_i(t),\omega_p)}$ like considered in \cite{vincent_effect_2004}, but in the present work we will discard this effect. Ants may also preferably choose the largest trails indicated by a large concentration of pheromones in one given direction. This 'preferential choice' will be discussed in connection to the kinetic and fluid models in section \ref{sec:kinetic} but discarded in the numerical simulations of the Individual-Based Model.

The frequency of the Poisson process is given by $\lambda_p M_i (t)$ where $M_i(t)$ is the number of pheromones in the detection region of Ant $i$: $M_i(t) = \mbox{Card} ( S_i(t) )$, and $\lambda_p$ is the trail recruitment frequency per unit pheromone. The dependency of the jump frequency upon the number of detected pheromones accounts for the observed increase of the alignment probability with the pheromone density. Of course, nonlinear functions of the pheromone density could be chosen as well. For instance, some saturation of the detection capability occurs at large pheromone densities, such as investigated in \cite{rauch_pattern_1995}. This effect will also be discarded here. We also discard any consideration of the detection mechanism, such as discussed e.g. in \cite{calenbuhr_model_1992,calenbuhr_model_1992-1, couzin_self-organized_2003}.

During their walk, ants leave trail pheromones at a certain deposition rate $\nu_d$. If at time $t$, ant $i$ deposits a pheromone, a new pheromone particle is created at position $x_i(t)$ with the direction $\omega_i(t)$. Hence, we postulate that:
\begin{displaymath}
  \mbox{At deposition times $t$, a pheromone } p \mbox{ is created with } (x_p,\omega_p) = (x_i(t), \omega_i(t)).
\end{displaymath}
Pheromones have a life-time $T_p$ and remain immobile during their lifetime. In this work, pheromone diffusion is neglected. Pheromone deposition and evaporation times are modeled by Poisson processes: each ant has a probability $\nu_d$ per unit of time to lay down a pheromone and each pheromone has a probability $1/T_p$ per unit of time to disappear.

Pheromone deposition mediates the interactions between the ants. This interaction is nonlocal in both space and time (because the ant which has deposited a pheromone may have moved away quite far before another ant interacts with it). Random velocity jumps and trail recruitment jumps have opposite effects. Random velocity jumps generate diffusion at large scales whereas trail recruitment jumps tend to produce concentrations of the ants trajectories on the pheromone trails. Therefore, the pheromone-meditated interaction induces correlations of the ants motions and these correlations result in trail formation.

The trail recruitment process together with pheromone evaporation result in network plasticity. To illustrate this mechanism, let us consider the following simplified situation.  Suppose an ant reaches a ``crossroads'' of trails, meaning a spot where pheromones point in two different directions denoted by $1$ and $2$. Suppose there are $n_1$ pheromones in one direction and $n_2$ in the other one.  The probability for the ant to choose to orient in direction $i$; $i = 1, 2$, is equal to the ratio $\frac{n_i}{n_1+n_2}$.  When the ant turns to its new direction, it may release a pheromone which will serve to reinforce this branch.  Eventually, one under-selected branch of the crossroads will vanish due to evaporation of the pheromones. Note that the choice of the surviving branch depends on random fluctuations of this process: therefore, the outcome of this situation is non-deterministic and even an initially strongly populated branch has a non-zero probability to vanish away.

\setcounter{equation}{0}
\section{Simulations and results}
\label{sec:simulation}

\subsection{Choice of the modeling and numerical parameters}
\label{subsec:choice}

We use experimentally determined parameter values as often as possible. Since parameters are species-dependent, we focus on the species 'Lasius Niger'.

In our model, the motion of a single ant is described by three quantities: the speed c, the frequency of random velocity jumps $\lambda_r$ and their amplitude $\sigma$. These three parameters have been estimated in different studies \cite{bernadou_does_2008,casellas_individual_2008} which give us a range of possible values. We choose rather low estimations of $\lambda_r$ and $\sigma$ (see Table \ref{table:parameters}) since in real experiments the estimation of these coefficients counts both for random jumps and recruitment by trails. The deposition rate of pheromones $\nu_d$ and their life time $T_p$ have also been measured experimentally for 'Lasius Niger' \cite{beckers_trail_1992}. After leaving a food source, an ant drops on the average $.5$ pheromone per second. This experimental value gives us an upper bound for $\nu_d$ because it corresponds to an estimation of $\nu_d$ in a very specific situation where the ant activity level is high. In our simulations, we use $\nu_d=.2\,s^{-1}$. Since in our model, all the ants lay down pheromones, we also take a low estimation of the pheromone lifetime ($T_p=100\,s$) otherwise the domain becomes saturated with pheromones.

By contrast, the interaction between ants and pheromones has not been quantified experimentally. For this reason, we do not have experimental values for the pheromone detection radius $R$ and the alignment probability per unit of time $\lambda_p$. In our simulations, we fix the radius of perception $R$ equal to $1$ cm (corresponding roughly to 2 body lengths). The alignment probability $\lambda_p$ remains a free parameter in our model. By changing the value of $\lambda_p$, we can tune the influence of the pheromone-mediated interaction between the ants. A low value of $\lambda_p$ corresponds to a weak interaction, whereas, for large values of $\lambda_p$, the ant velocities become controlled by the pheromone directions. In our simulations, $\lambda_p$ varies from $0$ to $3\,s^{-1}$.

For simplicity, all simulations are carried out in a square domain of size $L=100$ cm with periodic boundary conditions. For the initial condition, $200$ ants are randomly distributed in the domain. Their velocity $\omega_i$ is chosen uniformly on the circle ${\mathbb S}^1$. The ant-pheromone interaction is always taken nematic unless otherwise stated.

We can estimate the average number of pheromones $\langle M \rangle$ at equilibrium, when the average is taken over realizations. The evolution of $\langle M \rangle(t)$ is given by the following differential equation:
\begin{displaymath}
  \frac{d\langle M \rangle (t)}{dt} = \nu_d N - \frac{1}{T_p} \langle M \rangle (t),
\end{displaymath}
where $N$ is the number of ants, $\nu_d$ and $T_p$ are (resp.) the deposition rate and pheromone lifetime. Thus, at equilibrium (i.e. $d \langle M \rangle /dt =0$), the average number of pheromones is given by $\nu_d\,T_p\,N$. For our choice of parameters (Table \ref{table:parameters}), this corresponds to $4000$ pheromones.

\begin{table}
  \centering
  \begin{tabular}{ c | l | l }
    & {\bf Parameters} & {\bf Value} \\
    \hline
    $L$ & Box size  & 100 cm\\
    $N$ & Number of ants & 200\\
    \hline
    $c$ & Ant speed  & 2 cm/s\\
    $\lambda_r$ & Random jump frequency & 2 $s^{-1}$\\ 
    $\sigma$ & Random jump standard deviation & .1\\ 
    \hline
    $\nu_d$ & Pheromone deposition rate & .2 $s^{-1}$\\ 
    $T_p$ & Pheromone lifetime & 100 s\\       
    \hline
    $R$ & Detection radius & 1 cm\\
    $\lambda_p$ & Trail recruitment frequency & 0-3 $s^{-1}$
  \end{tabular}
  \caption{Table of the parameters used in the simulations.}
  \label{table:parameters}
\end{table}

\subsection{Detection of trails}
\label{subsec:detection}

\subsubsection{Evidence of trail formation}
\label{subsubsec:trailform}

The typical outcomes of the model are shown in figure \ref{fig:illustration_trail}. After an initial transient, we observe the formation of a network of trails. This network is not static, as we observe in the two graphics: the network at time $t=2000\,s$ is significantly different from the network observed at time $t=1000\,s$. Here, the goal is to provide statistical descriptors of this trail formation phenomenon and to analyze it.

\begin{figure}[ht]
  \centering
  \includegraphics[scale=.5]{ant_N200_NP_3988_t1000.eps} \quad 
  \includegraphics[scale=.5]{ant_N200_NP_3988_t2000.eps}
  \caption{A typical output of the model at two different times. The ants are represented in blue and the pheromones in green. We clearly observe the formation of trails. Parameters of the simulation: $\lambda_p=2 s^{-1}$, $\Delta t=.05$ (see also Table \ref{table:parameters}).}
  \label{fig:illustration_trail}
\end{figure}

\subsubsection{Definition of a trail}
\label{subsubsec:traildef}

\begin{figure}[p]
  \centering
  \includegraphics[scale=.5]{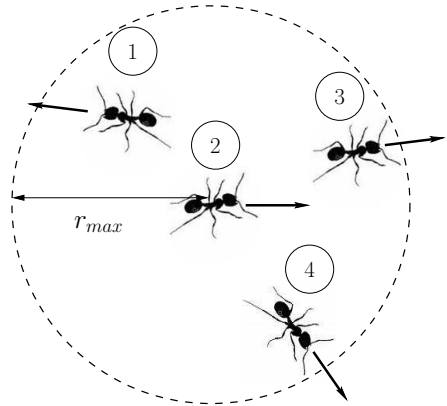}
  \caption{In this example, Ants 1, 2 and 3 are linked together: they form a {\it trail}. Ant 4 is not linked to Ant 2 since their directions are too different.}
  \label{fig:illustration_connexion}
\end{figure}

To quantify the amount of particles that are organized into trails at a given time, we consider the collection of all particles, that is to say, the union of the sets of ants and of pheromones. Indeed collecting the pheromones allows us to trace back the recent history of the individuals. To define a trail, we fix two parameters: a distance $r_{\max}$ and an angle $\theta_{max}$. We say that particle $P_i=(x_i,\omega_i)$ ($P_i$ being either an ant or a pheromone) is {\it linked} to particle $P_j=(x_j,\omega_j)$ if the distance between the two particles is less than $r_{max}$ and the angle between $\omega_i$ and $\omega_j$ is either less than $\theta_{max}$ or greater than $\pi-\theta_{max}$. In other words, we define a relationship (see figure \ref{fig:illustration_connexion}):
\begin{equation}
  \label{eq:relationship}
  P_i \sim P_j \quad \text{ if and only if } \quad |x_i-x_j|<r_{max} \;\text{ and }\; \sin
  (\omega_j-\omega_i) < \sin \theta_{max}.
\end{equation}
Using this relationship, particles can be sorted into different trails: we say that $P$ and $Q$ belong to the same {\it trail} if there exists particles $P_1, \ldots, P_k$ ({\it a path}) such that $P \sim P_1$, $P_1 \sim P_2, \ldots$, $P_k \sim Q$. Thus, a trail is defined as the connected components of the particles under the relationship (\ref{eq:relationship}). A trail is approximately a slowly turning lane of particles.

\begin{figure}[p]
  \centering
  \includegraphics[scale=.6]{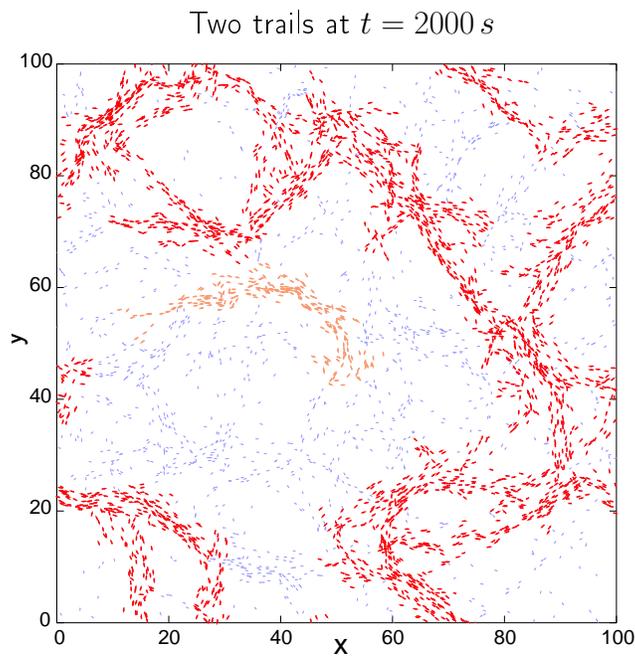}
  \caption{The two largest trails (drawn in red and orange) for the simulation depicted in figure \ref{fig:illustration_trail} at time $t=2000\,s$. Parameters for the estimation of the trails: $r_{max}=2$ cm and $\theta_{max} = 45^\circ$.}
  \label{fig:CC_N200_T2000_r2_theta45}
\end{figure}

As a first example, in figure \ref{fig:CC_N200_T2000_r2_theta45}, we display the partitioning into trails of the previous simulations (figure \ref{fig:illustration_trail}) at time $t=2000\,s$ with $r_{max}=2$ cm and $\theta_{max} = 45^\circ$. For these values, the largest trail (drawn in red in figure \ref{fig:CC_N200_T2000_r2_theta45}) consists of 2670 particles and the second largest (drawn in orange in figure \ref{fig:CC_N200_T2000_r2_theta45}) is made of 254 particles.

\subsubsection{Statistics of the trails}
\label{subsubsec:statistics}

We expect that trail formation results in the development of a small number of large trails, while unorganized states are characterized by a large number of small trails, most of them being reduced to single elements. Therefore, trail formation can be detected by observing the trail sizes. With this aim, we denote by $S_i(t)$ the size of the trail to which particle $i$ belongs at time $t$. Let ${\mathcal N}(t)$ be the total number of particles, i.e. ${\mathcal N}(t) = N + {\mathcal P}(t)$ where $N$ is the number of ants and ${\mathcal P}(t)$ is the number of pheromones at time $t$. We form
\begin{displaymath}
  p_t(S) = \frac{ \mbox{Card} ( \{i \, | \, S_i(t) = S \} ) }{{\mathcal N}(t)}, \quad S \in {\mathbb N}.
\end{displaymath}
$p_t(S)$ is the probability that a particle belongs to a trail of size $S$ at time $t$. An unorganized state is therefore characterized by a quickly decaying $p_t(S)$ as a function of $S$ while a state where the particles are highly organized into trails displays a bimodal $p_t(S)$ with high values for large values of $S$. To display the distribution of $p_t(S)$ is easy: it is nothing but the histogram of the trail sizes $S_i$, collected from several independent simulations with identical parameters.

As an illustration, we provide the distribution $p_t(S)$ for the set parameters used to generate Figs. \ref{fig:illustration_trail} and \ref{fig:CC_N200_T2000_r2_theta45}, (i.e. $\lambda_p=2\,s^{-1}$, $r_{max}=2$ cm and $\theta_{max} = 45^\circ$), with $1000$ realizations. We clearly observe in Fig.  \ref{fig:density_cluster_r2_angle45} (left) that the distribution $p_t(S)$ is bimodal: a first maximum is observed near the minimal value of $S$, i.e.  $S=1$, and a second maximum is observed near the values $S \approx 2500$. This indicates that a particle (i.e. an ant or a pheromone) belongs to either a small-size trail ($S<100$) or to a large-size trail ($S \approx 2500$). As a control sample for our statistical measurement, we run the same simulations but cutting off the ant-pheromone interaction (i.e. $\lambda_p=0$) and proceed to the same analysis. In Fig. \ref{fig:density_cluster_r2_angle45} (right), we observe that without the influence of the pheromones (blue histogram) the probability $p_t(s)$ is only concentrated near the value $S=1$ and decays very fast to almost vanish for $S > 500$.

\begin{figure}[ht]
  \centering
  \includegraphics[scale=.4]{density_cluster_r2_angle45_nSimu1000.eps} \hspace{.1cm}
  \includegraphics[scale=.4]{density_cluster_r2_angle45_nSimu1000_dos}
  \caption{ ({\bf Left}) Histogram of the trail sizes $S$ estimated from $1000$ realizations. ({\bf Right}) Histograms of $S$ with and without trail-pheromone interaction ($\lambda_p=2$ and $\lambda_p=0$ resp.). The parameters for this simulation are the same as in Figs.  \ref{fig:illustration_trail} and \ref{fig:CC_N200_T2000_r2_theta45}.}
  \label{fig:density_cluster_r2_angle45}
\end{figure}

\subsection{Trail size}
\label{subsec:length}

As observed in figure \ref{fig:density_cluster_r2_angle45}, ant-pheromone interactions lead to the formation of trails which are evidenced by the transformation of the shape of the distribution $p_t(S)$. To perform a systematic parametric analysis of the trail formation phenomenon, we use the mean $\langle S\rangle$ of the distribution $S$:
\begin{displaymath}
  \langle S\rangle = \sum_{S \in {\mathbb N}} S \, p_t(S).
\end{displaymath}
The quantity $ \langle S\rangle$ quantifies the level of organization of the system into trails. Indeed, large values of $\langle S\rangle$ indicate a high level of organization into trails while smaller values of $\langle S\rangle$ are the signature of a disordered system. For example, in Fig. \ref{fig:density_cluster_r2_angle45}, we have $\langle S\rangle = 1333.7$ \, when the ant-pheromone interaction is on with interaction frequency $\lambda_p=2$. By contrast, its value falls down to $\langle S\rangle = 76.8$ when the ant-pheromone interaction is turned off (i.e. $\lambda_p=0$) and the system is in a fully disordered state.

\begin{figure}[ht]
  \centering
  \includegraphics[scale=.55]{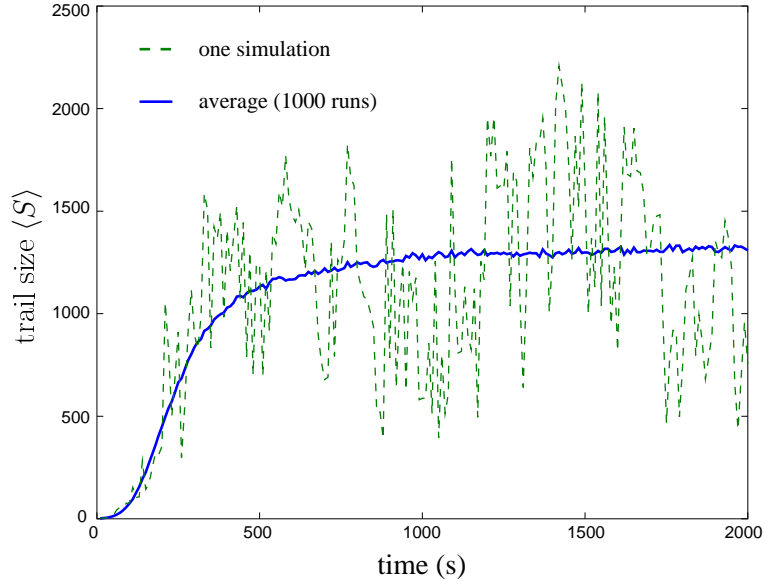}
  \caption{ Mean trail size $ \langle S\rangle $ as a function of time for one simulation (dashed line) and averaged over $1000$ different simulations (solid line).}
  \label{fig:S(t)}
\end{figure}

Our first use of the mean trail size $ \langle S\rangle $ is to show that it stabilizes to a fixed value after an initial transient. Fig \ref{fig:S(t)} shows the mean trail size $ \langle S\rangle (t)$ for one simulation (dashed line) and averaged over $1000$ different simulations (solid line). It appears that,  after some transient, $ \langle S\rangle (t)$ presents a lot of fluctuations about an averaged value. If the simulation is reproduced a large number of times and the mean trail size $ \langle S\rangle (t)$ is averaged over all these realizations, the convergence towards a constant value becomes apparent. 

Therefore, statistical analysis of the trail patterns using the mean trail size $ \langle S\rangle $ become significant only once this constant value has been reached. In the forthcoming sections, analysis will be performed for simulation times equal to $2000$ s, which is significantly larger than the time needed for the stabilization of  $ \langle S\rangle $ (about $800$ s). 

\subsection{Evidence of a phase transition}
\label{subsec:evidence}

\begin{figure}[p]
  \centering
  \includegraphics[scale=.5]{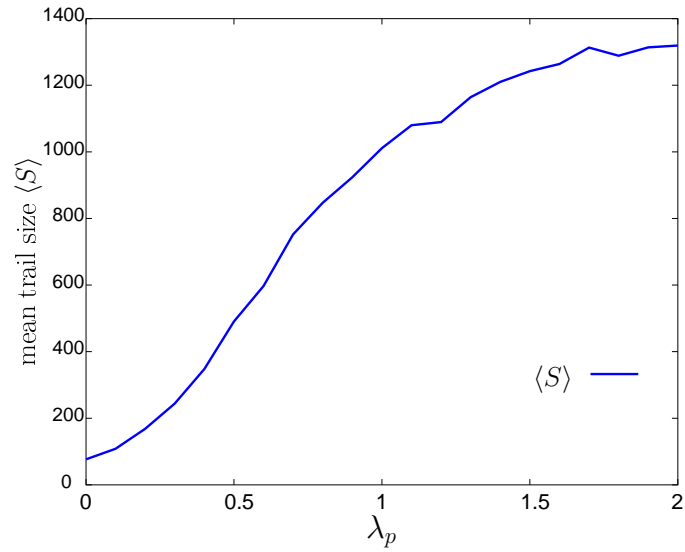}
  \caption{The mean $\langle S\rangle$ of the distribution $p_t(s)$ as a function of the ant-pheromone interaction frequency $\lambda_p$ for a fixed value of the random jump frequency $\lambda_r$.}
  \label{fig:meanS_r2_angle45}
\end{figure}

In Fig. \ref{fig:meanS_r2_angle45}, we display $\langle S\rangle$ as a function of $\lambda_p$. For each value of $\lambda_p$, we estimate $\langle S\rangle$ by averaging it over $1000$ independent simulations. We observe an abrupt increase of $\langle S\rangle$ when $\lambda_p$ varies from $0$ to $1$ which means that a sharp transition from an unorganized system to a system organized into trails arises. For larger values of $\lambda_p$, the influence of the ant-pheromone interaction saturates and $\langle S\rangle$ reaches a plateau at the approximate value $\langle S\rangle \approx 1300$.

\begin{figure}[p]
  \centering
  \includegraphics[scale=.5]{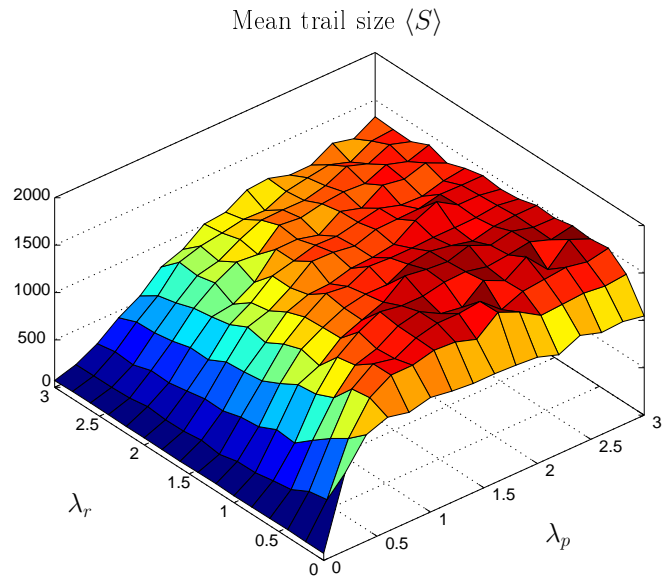}
  \caption{The mean $\langle S\rangle$ of the distribution $p_t(s)$ as a function of the pair $(\lambda_r, \lambda_p)$, The cuts of this surface at a fixed value of $\lambda_r$ shows the same behavior as in figure \ref{fig:meanS_r2_angle45}.}
  \label{fig:meanS3D_r2_angle45}
\end{figure}

The transition from disorder to trails also depends on the other parameters of the model. For example, if we increase the noise by increasing the random jump frequency $\lambda_r$, the corresponding value of $\langle S\rangle$ decreases. In order to restore the previous value of $\langle S\rangle$ the ant-pheromone interaction frequency $\lambda_p$ must be increased simultaneously. In Fig.  \ref{fig:meanS3D_r2_angle45}, we plot $\langle S\rangle$ as a function of both the random jump frequency $\lambda_r$ and the ant-pheromone interaction frequency $\lambda_p$. We estimate $\langle S\rangle$ by averaging it over $100$ realizations for each value of the pair $(\lambda_p, \lambda_r)$. We still observe a fast transition from an unorganized state ($\langle S\rangle<100$) to a state organized into trails ($\langle S\rangle\geq1000$) when $\lambda_p$ increases. However, as the noise $\lambda_r$ increases, this transition becomes smoother. Moreover, the plateau reached by $\langle S\rangle$ when $\lambda_p$ is large is still comprised between $1300$ and $1500$ for all values of $\lambda_r$, but reaching this plateau for large values of $\lambda_r$ requires larger value of $\lambda_p$.

At the value $\lambda_r=0$, the transition from disorder ($\langle S\rangle\leq 100$) to trail-like organization ($\langle S\rangle \geq 1000$) is the fastest. However, the plateau reached by  $\langle S\rangle$ when $\lambda_p$ is large is significantly lower than for larger values of $\lambda_r$ ($\langle S\rangle \approx 1000$ instead of $1300$). This could be attributed to the fact that, without random jumps, the level of diffusion is too low, the ants do not mix enough, and trails have little opportunities to merge. 

On the other hand, we can look for another explanation of this paradoxical lower value of $\langle S\rangle$ when $\lambda_r$ is very small. Indeed, we notice that, in this case, the formed trails are much narrower than for larger values of $\lambda_r$. Fig. \ref{fig:simuLowLdR} (left) shows a simulation result using a quite small random jump frequency of $\lambda_r=.2$. We observe that the trails are narrower and more straight than those obtained with the larger value $\lambda_r=2$ (figure \ref{fig:illustration_trail}). We can quantify statistically this feature by changing the parameters of trail detection $r_{max}$ and $\theta_{max}$. We reduce the maximum distance ($r_{max}=1.5$ cm) and the maximum angle ($\theta_{max}=35^\circ$). With these smaller values, two particles are less likely to be connected. Then we proceed to the same analysis as in figure \ref{fig:meanS3D_r2_angle45}, by estimating the mean size of the trails $\langle S\rangle$ as a function of $\lambda_r$ and $\lambda_p$, averaged over $100$ realizations. As we observe in figure \ref{fig:simuLowLdR} (right), the mean size of the trails $\langle S\rangle$ is much larger for smaller values of $\lambda_r$ and we recover the same behavior as that observed for larger values of $\lambda_r$. This discussion illustrates the difficulty of working with an estimator which depends on arbitrary choices of scales (here the space and angular threshold of trail detection). A discussion of the dependence of the trail width upon the biological parameters is developed in the next section.

\begin{figure}[ht]
  \centering
  \includegraphics[scale=.4]{simu2_ldP2_ldR02.eps} \qquad 
  \includegraphics[scale=.4]{meanS3D_r15_angle35.eps}
  \caption{({\bf Left}) A simulation with low noise ($\lambda_r=.2$); the other parameters are the same as in figure \ref{fig:illustration_trail} ($t=2000\,s$). ({\bf Right}) The mean size of the trails $\langle S\rangle$ estimated with $r_{max}=1.5$ cm and $\theta_{max}=35^\circ$.}
  \label{fig:simuLowLdR}
\end{figure}

\subsection{Trail width}
\label{subsec:width}

A way to highlight the dependence of the trail width upon the model parameters is to compute a two-particle correlation distribution. Let a particle (ant or pheromone) $i$ be located at position $x_i$ and velocity $\omega_i$. Denote by $\omega_i^\bot$ the orthogonal vector to $\omega_i$ in the direct orientation. For all particles $j \not = i$, we form the vector
\begin{displaymath}
  X_{ij} = \left( \begin{array}{c} (x_j - x_i) \cdot \omega_i^\bot \\ (x_j - x_i) \cdot \omega_i \end{array} \right).
\end{displaymath}
The distribution $\frac{2}{{\mathcal N}({\mathcal N}-1)} f_2(X)$, with 
\begin{displaymath}
  f_2(X) = f_2(X_x, X_y) =  \sum_{(i,j), \, i \not = j} \delta(X-X_{ij}),
\end{displaymath}
where $\delta$ is the Dirac delta, provides the probability that, given a first particle (located at say $x_0$ with orientation $\omega_0$), a second particle lies at location $x_0 + \omega_0^\bot X_x + \omega_0 X_y$ (see figure \ref{fig:two-part} (left) for an illustration of the construction of $f_2$). Looking at this 2-particle density, trails appear as concentrations near a line passing through the origin and directed in the $y$-direction. Figure \ref{fig:two-part} (right) provides a histogram of the two-particle density $ f_2$ for the simulation corresponding to the right picture of fig. \ref{fig:illustration_trail}. The above mentioned concentration is clearly visible. Additionally, the typical width of this concentration gives access to the typical width of the trails.

\begin{figure}[p]
  \centering
  \includegraphics[width=6.5cm]{estimationF2.eps} \quad
  \includegraphics[width=7.5cm]{f2_2D.eps} 
  \caption{({\bf Left}) Construction of the two-particle distribution $f_2$. ({\bf Right}) Histogram of the two-particle density $f_2$ for the test-case corresponding to the right picture of fig. \ref{fig:illustration_trail}. $f_2(X)$ is represented via a color scale as a function of the two components of $X$. }
  \label{fig:two-part}
\end{figure}

In order to better estimate the typical width of the trails, we plot cuts of the two-particle density $f_2$ along the line $\{y = 0\}$ (see figure \ref{fig:two-part}). In practice, these cuts are determined by computing the following density
\begin{displaymath}
  \bar f_2(r) =   \sum_{(i,j), \, i \not = j, \, |(X_{ij})_y| \leq \xi } \delta(r-|(X_{ij})_x|).
\end{displaymath}
where $\xi$ is suitable chosen (of the order of $1$ cm). Figure \ref{fig:two-part_cuts} (left) displays $\bar f_2(r)$ as a function of $r$ for different values of the trail recruitment frequency $\lambda_p$ and a fixed value of the random jump frequency equal to $\lambda_r = 2$ s$^{-1}$. It appears that $\bar f_2$ is higher and decreases faster for larger values of $\lambda_p$. The decay of $\bar f_2$ can give an estimate of the width of the trail: if we approximate the decay of $\bar f_2$ by an exponential,
\begin{displaymath}
  \bar f_2(r) \approx f_0\, \exp\left(-\frac{r}{r_0}\right) \qquad  \text{for} \quad r \approx 0,
\end{displaymath}
then $r_0$ measures the typical width of the trail. This quantity can be estimated using the formula:
\begin{equation}
  \label{eq:formula_r0}
  r_0 = \frac{1}{|(\ln \bar f_2)'(0)|}.
\end{equation}
As we observe in Table \ref{table:estimation_r0}, the width $r_0$ increases as $\lambda_p$ decreases. Therefore, increasing the trail recruitment frequency increases the intensity of the particles interactions and produces trails with smaller width. Figure \ref{fig:two-part_cuts} (right) displays $\bar f_2(r)$ as a function of $r$ for different values of the random jump frequency $\lambda_r$ and a fixed value of the trail recruitment frequency equal to $\lambda_p = 2$ s$^{-1}$. Here, the trail width $r_0$ estimated from $\bar f_2$ is larger for large values of $\lambda_r$ (see Table \ref{table:estimation_r0}), indicating that the typical width of the trails increases with increasing $\lambda_r$, as it should.

We also observe a discontinuity at $r=0$ for all the functions $\bar f_2$ (figure \ref{fig:two-part_cuts}). These jumps are easily explained by the deposit process: each time an ant drops a pheromone, the new pheromone and the ant are located at the same position exactly. This results in a peak of concentration of $\bar f_2$ at $r=0$.


\begin{figure}[p]
  \centering
  \includegraphics[width=6.5cm]{f2_radial_ldP.eps} \quad 
  \includegraphics[width=6.5cm]{f2_radial_ldR.eps}
  \caption{({\bf Left}) $\bar f_2(r)$ as a function of $r$ for different values of the trail recruitment frequency $\lambda_p$  and a fixed value of the random jump frequency equal to $\lambda_r = 2$ s$^{-1}$. ({\bf Right}) $\bar f_2(r)$ as a function of $r$ for different values of the random jump frequency $\lambda_r$ and a fixed value of the trail recruitment frequency equal to $\lambda_p = 2$ s$^{-1}$.}
  \label{fig:two-part_cuts}
\end{figure}

\begin{table}[ht]
  \centering
  \begin{tabular}[c]{l|c}
    $\lambda_p$ & $r_0$ (cm)\\
    \hline
    3 & 2.796 \\
    2 & 3.181 \\
    1 & 4.350 \\
    &
  \end{tabular} \qquad \qquad
  \begin{tabular}[c]{l|c}
    $\lambda_r$ & $r_0$ (cm) \\
    \hline
    0 & 2.532 \\
    1 & 2.964 \\
    2 & 3.181 \\
    3 & 3.345
  \end{tabular}
  \caption{Estimations of the width $r_0$ (\ref{eq:formula_r0}) of the trails using $\bar f_2$ given in figure \ref{fig:two-part_cuts}. We estimate the derivative of $\ln \bar f_2(r)$ near $0$ using the values of $r$ between $.1$ and $2$.}
  \label{table:estimation_r0}
\end{table}


\setcounter{equation}{0}
\section{Kinetic and continuum descriptions}
\label{sec:kinetic}


\subsection{Framework}
\label{subsec:framework}

In this section, we propose meso- and macro-scopic descriptions of the previously discussed ant dynamics. We first propose a kinetic model, i.e. a model for the probability distributions of ants and pheromones. The derivation of this kinetic model is formal and based on analogies with the underlying discrete dynamics. A rigorous derivation of the kinetic model from the discrete dynamics is up to now beyond reach. Issues such as the validity of the chaos propagation property \cite{cercignani_mathematical_1994}, which is the key for proving such results, may be quite difficult to solve. Then, fluid limits of this kinetic model will be considered. We will notice that the resulting fluid models can only exhibit the development of trails if some concentration mechanism is added, while the numerical simulations above indicate that such a mechanism is not needed at the level of the Individual-Based Model.


\subsection{Kinetic model}
\label{subsec:kinetic}

In this section, we introduce the kinetic model of the discrete ant-pheromone interaction on a purely formal basis.  We introduce the ant distribution function $F(x, \omega, t)$ and the pheromone distribution function $G(x, \omega, t)$, for $x \in {\mathbb R}^2$, $\omega \in {\mathbb S}^1$ and $t\geq 0$. They are respectively the number density in phase-space $(x,\omega)$ of the ants (respectively of the pheromones), i.e. the number of such particles located at position $x$ with orientation $\omega$ at time $t$. Here, we remark that the consideration of directed pheromones requires the introduction of a pheromone density in (position, orientation) phase-space in the same manner as for the ants.

\medskip
\noindent
{\bf Trail dynamics.} The trail dynamics is described by the ordinary differential equation:
\begin{equation}
  \partial_t G(x, \omega, t) = \nu_d F(x, \omega, t) - \nu_e G(x, \omega, t), 
  \label{eq:kinetic_phero}
\end{equation}
This equation can be easily deduced from the evolution of the probability density of the underlying stochastic Poisson process. The first term describes deposition by the ants according to a Poisson process of frequency $\nu_d$ while the second term results from the finite lifetime expectancy $T_p = \nu_e^{-1}$ of the pheromones. Pheromones are supposed immobile, which explains the absence of any convection or diffusion operator in this model. Discarding pheromone diffusion is done for simplicity only and can be easily added. It would add a term $\Delta_x G$ or $\Delta_\omega G$ at the right-hand side of (\ref{eq:kinetic_phero}) according to whether one considers spatial or orientational diffusion. 

\medskip
\noindent
{\bf Ant dynamics.} The evolution of the ant distribution function is ruled by the following kinetic equation:
\begin{equation}
  \partial_t F + c \, \omega \cdot \nabla_x F = Q(F).
  \label{eq:kinetic_ants}
\end{equation}
The left-hand side describes the ant motion with constant speed $c$ in the direction $\omega$. The right hand side is a Boltzmann-type operator which describes the rate of change of the distribution function due to the velocity jump processes. $Q$ is decomposed into
\begin{displaymath}
  Q = Q_r + Q_p,
\end{displaymath}
where $Q_r$ and $Q_p$ respectively describe the random velocity jumps and the trail recruitment jumps.  

Both operators $Q_k(x,\omega, t)$, $k=p$ or $r$ express the balance between gain and loss due to velocity jumps, i.e. $Q_k(x,\omega, t) = Q_k^+ - Q_k^-$. The gain term $Q_k^+$ describes the rate of increase of $F(x,\omega,t)$ due to particles which have post-jump velocity $\omega$ and pre-jump velocity $\omega'$. Similarly, the loss term $Q_k^-$ describes the rate of decay of $F(x,\omega,t)$ due to particles jumping from $\omega$ to another velocity $\omega'$. The jump probability $P_k(\omega \rightarrow \omega') d \omega'$ is the probability per unit time that a particle with velocity $\omega$ jumps to the neighborhood $d \omega'$ of $\omega'$ due to jump process $k$. Therefore, the expression of $Q_k$ is: 
\begin{equation}
  Q_k(F) (x,\omega,t) = \int_{\mathbb{S}^1} \big( P_k(\omega' \rightarrow \omega) F(x, \omega', t) - P_k(\omega \rightarrow \omega') F(x, \omega, t) \big) d \omega'. 
  \label{eq:Qk}
\end{equation}
where the positive term corresponds to gain and the second term, to loss. By symmetry, we note that 
\begin{equation}
  \int_{\mathbb{S}^1} Q_k(F) (x,\omega,t) \, d \omega = 0,
  \label{eq:int_Qk=0}
\end{equation}
for any distribution $F$. This expresses that the local number density of particles is preserved by the velocity jump process. 

Now we describe the expressions of the jump probabilities $P_k$. For both processes, we postulate the existence of a detailed balance principle, which means that the ratio of the direct and inverse collision probabilities are equal to the ratios of the corresponding equilibrium probabilities
\begin{equation} 
  \frac{P_k(\omega' \rightarrow \omega)}{P_k(\omega \rightarrow \omega')} = \frac{h_k(\omega)}{h_k(\omega')} , 
  \label{eq:detailed_balance}
\end{equation}
where $h_k$ is the equilibrium probability of the process $k$ ($k=r$ or $k=p$). 
Using (\ref{eq:detailed_balance}), we can define:
\begin{displaymath}
  \Phi_k(\omega',\omega)= \frac{1}{h_k(\omega)} P_k(\omega' \rightarrow \omega) = \Phi_k(\omega,\omega'),
\end{displaymath}
which is symmetric by exchange of $\omega$ and $\omega'$ and write 
\begin{equation}
  Q_k(F) (x,\omega,t) = \int_{\mathbb{S}^1} \Phi_k(\omega,\omega') \big( h_k(\omega) \, F(x, \omega', t) - h_k(\omega') \, F(x, \omega, t)\big) d \omega'. 
  \label{eq:Qk2}
\end{equation}
From this equation, it is classically deduced that the equilibria, i.e. the solutions of $Q_k(F)=0$ are given by $F(x,\omega,t) = \rho(x,t) h_k(\omega)$ with arbitrary $\rho$. We recall the argument here for the sake of completeness. Indeed, such $F$ are clearly equilibria. Reciprocally, if $F$ is an equilibrium, then, using the symmetry of $\Phi_k$ leads to  
\begin{eqnarray*} 
  0 &=& \int_{\mathbb{S}^1}  Q_k(F) \, \frac{F}{h_k} \,  d\omega \\
  &=& - \frac{1}{2} \int_{({\mathbb{S}^1})^2} \Phi_k(\omega,\omega') \,  h_k(\omega) \,  h_k(\omega') \, \left( \frac{F(x, \omega', t)}{h_k(\omega')} - \frac{F(x, \omega, t)}{h_k(\omega)}  \right)^2 \, d \omega \, d \omega'. 
\end{eqnarray*} 
The last expression is the integral of a non-negative function which therefore must be identically zero for any choice of $(\omega,\omega')$. It follows that the only equilibria are functions of the form $\rho h_k$ with $\rho$ only depending on $(x,t)$. It is not clear if the biological processes actually do satisfy the detailed balance property but this hypothesis simplifies the discussion. Indeed, with this assumption, the equilibria $h_k$ and the jump probabilities $\Phi_k$ can be specified independently.

\medskip
\noindent
{\bf Trail recruitment jumps.} For trail recruitment, we first need to specify the equilibrium distribution as a function of the pheromone distribution. Several options are possible: non-local interactions, local ones, preferential choice, nematic interactions.

\medskip
\noindent
{\em 1.  Non-local interaction.} We first introduce the sensing application:
\begin{equation*}
  S_R(x, \omega, t) = \frac{1}{\pi R^2} \int_{|x-y|<R} G(y, \omega, t) d y,
\end{equation*}
where $R$ represents the perception radius of the particle, i.e. the maximal distance at which it can feel a deposited pheromones. The quantity $S_R(x, \omega, t)$ represents the density of pheromones pointing towards $\omega$ which can be perceived by an ant at point $x$ in its perception area. We also define
\begin{equation*}
  T_R(x, t) = \int_{\mathbb{S}^1} S_R(x, \omega, t) \, d \omega,
\end{equation*}
the pheromone total density within the perception radius, regardless of orientation. Then, we let the equilibrium distribution of the trail recruitment process as follows: 
\begin{equation}
  h_p(\omega) = g_R(x,\omega,t) :=  \frac{S_R(x,\omega,t)}{T_R(x,t)} , 
  \label{eq:equi_nonlocal}
\end{equation}
which, by construction, is a probability density. Now, The expression for the transition probability reads:
\begin{equation}
  \Phi_p (\omega \rightarrow \omega'; x,t) = \lambda_p \gamma (T_R(x,t)) \phi_p(\omega \cdot \omega'),
  \label{eq:proba_recruit}
\end{equation}
where $\lambda_p$ is the trail-recruitment frequency and $\gamma$ is a dimensionless increasing function of $T$ which accounts for the fact that recruitment by trails increases with pheromone density (in the discrete particle dynamics, we have taken $\gamma(T) = \pi R^2 T$, the total number of pheromones in the sensing region).  The function $\phi_p(\omega \cdot \omega')$ represents the angular dependence of the interaction process and is such that
\begin{equation} 
  \frac{1}{2 \pi} \int_{{\mathbb S}^1} \phi_p(\omega \cdot \omega') \, d \omega' = 1. 
  \label{eq:transition_normalization}
\end{equation}
We assume that it is independent of the pheromone distribution for simplicity. Inserting (\ref{eq:proba_recruit}) into (\ref{eq:Qk2}), the trail-recruitment operator is written:
\begin{eqnarray}
  & & \hspace{-1cm}
  Q_p(F) (x,\omega,t) = \lambda_p \gamma (T_R(x,t)) \int_{\mathbb{S}^1} \phi_p(\omega \cdot \omega') ( g_R(\omega) \, F(x, \omega', t)  \nonumber \\
  & & \hspace{7cm}
  - g_R(\omega') \, F(x, \omega, t)) d \omega'. 
  \label{eq:Qp}
\end{eqnarray}
The choice of $\phi_p$ which corresponds to the discrete dynamics discussed in the previous sections is $ \phi_p(\omega \cdot \omega') = 1$.  Inserting this prescription into (\ref{eq:Qp}) and using (\ref{eq:equi_nonlocal}) leads to the simplified operator
\begin{equation*}
  Q_p(F) (x,\omega, t) = \lambda_p \gamma (T_R(x,t)) \left( \rho(x, t)\frac{S_R(x,\omega,t)}{T_R(x,t)} - F(x, \omega, t) \right), 
\end{equation*}
with 
\begin{displaymath}
  \rho(x, t) = \int F(x, \omega, t) d \omega,
\end{displaymath}
the local ant density at $x$. 

\medskip
\noindent
{\em 2.  Local interaction.} This corresponds to taking the limit of the sensing radius to zero: $R \to 0$ which leads to 
\begin{displaymath}
  h_p(\omega) = g(\omega):= \frac{G(x,\omega,t)}{T(x,t)}, \quad T(x,t) = \int_{{\mathbb S}^1} G(x,\omega,t) \, d \omega.
\end{displaymath}
$T$ is the local trail density. Then, the expression of the collision operator is easily deduced from (\ref{eq:Qp}) by changing $g_R$ into $g$. In the case where $\phi_p = 1$, we get the expression:
\begin{equation*}
  Q_p(F) (x,\omega, t) = \lambda_p \gamma (T(x,t)) \left( \rho(x, t) g(x,\omega,t) - F(x, \omega, t) \right).
\end{equation*}

\medskip
\noindent
{\em 3.  Preferential choice.} We can envision a mechanism by which the ants can sense and choose the most frequently used trails. A possible way to model this preferential choice is by postulating an equilibrium distribution of the form 
\begin{equation}
  h_p(\omega) = g_R^{[k]} (\omega) = \frac{g_R^k(\omega)}{ \int_{{\mathbb S}^1} g_R^k(\omega) \, d \omega} , 
  \label{eq:(k)operation}
\end{equation}
with a power $k>1$. Indeed, it can be shown \cite{boissard_degond} that the maxima of $g_R^{[k]}$ are larger than those of $g_R$ and similarly, the minima are lower. Additionally, the monotony is preserved, i.e. 
\begin{displaymath}
  g_R(\omega) \leq g_R(\omega') \Longrightarrow g_R^{[k]}(\omega) \leq g_R^{[k]}(\omega'), \quad \forall (\omega,\omega') \in ({\mathbb S}^1)^2.
\end{displaymath}
Therefore, taking $g_R^{[k]} (\omega)$ as equilibrium distribution of the ant-pheromone interaction means that the ants choose the trails $\omega$ with a higher probability when the trail density in direction $\omega$ is high and with lower probability when the trail density is low. The expression of the collision operator is easily deduced from (\ref{eq:Qp}) by changing $g_R$ into $g_R^{[k]}$. In the case where $\phi_p = 1$, we get:
\begin{equation*}
  Q_p(F) (x,\omega, t) = \lambda_p \gamma (T_R(x,t)) \left( \rho(x, t) g_R^{[k]}(x,\omega,t) - F(x, \omega, t) \right).
\end{equation*}
This mechanism can also be combined with a local interaction, by replacing $g_R$ by the local angular pheromone probability $g$. We note that this mechanism is not implementable in the discrete dynamics because the operation $g \to g^{[k]}$ is only defined for measures $g$ which belong to the Lebesgue space $L^k({\mathbb S}^1)$. However, sums of Dirac deltas, which correspond to the measure $g$ in the Individual-Based Model, do not belong to this space. Therefore, a smoothing procedure must be applied to such measures beforehand. Since, it is not possible to obtain experimental data about the smoothing procedure and the power $k$, the preferential choice model has not been used in the numerical experiments of the previous sections. 

\medskip
\noindent
{\em 4.  Nematic interaction.} The above described ant-pheromone interactions are polar ones, i.e. the pheromones are supposed to have both a direction and an orientation. However, we can easily propose a nematic interaction, for which an ant of velocity $\omega$ chooses $\omega'$ among the pheromone directions and their opposite in such a way that the angle $\widehat{(\omega,\omega')}$ is acute, i.e. such that $\omega \cdot \omega' > 0$. For this purpose, we modify the equilibria of the trail recruitment operator as follows: 
\begin{displaymath}
  h_p(x,\omega,t) =  g_R^{(sym)} := \frac{S_R(x,\omega,t) + S_R(x,-\omega,t)}{2T_R(x,t)},
\end{displaymath}
and suppose that 
\begin{equation} 
  \phi_p(\omega \cdot \omega') = 0, \quad \mbox{ when } \omega \cdot \omega' \leq 0. 
  \label{eq:restrict}
\end{equation}
The expression of the collision operator is easily deduced from (\ref{eq:Qp}) by making the change of $g_R$ into $g_R^{(sym)}$ and imposing the restriction (\ref{eq:restrict}). 
In the case where 
\begin{displaymath}
  \phi_p(\omega \cdot \omega') = 2 H(\omega \cdot \omega'),
\end{displaymath}
where $H$ is the Heaviside function (i.e. the indicator function of the positive real line), we find 
\begin{eqnarray*}
  & & \hspace{-1cm}
  Q_p(F) (x,\omega, t) = \lambda_p \gamma (T_R(x,t)) \left[ \frac{1}{T_R(x,t)} \left( \rho_\omega^+ (x, t) \, S_R(x,\omega,t) +  \right.  \right.\\ 
  & & \hspace{5cm}
  \left. \phantom{\frac{1}{T_R(x,t)}} \left. + \rho_{\omega}^- (x, t) \, S_R(x,- \omega,t) \right) 
    - F(x, \omega, t) \right], 
\end{eqnarray*}
with 
\begin{displaymath}
  \rho_{\omega}^\pm(x, t) = \int F(x, \omega', t) \, H(\pm \omega \cdot \omega') \,  d \omega',
\end{displaymath}
is the local density of ants pointing in a direction making respectively an acute angle (for $\rho_{\omega}^+$) or obtuse angle (for $\rho_{\omega}^-$) with $\omega$ at $x$.

\medskip
\noindent
{\bf Random velocity jumps.} For random velocity jumps, we assume a uniform equilibrium 
\begin{equation*} 
  h_r(\omega) = \frac{1}{2 \pi} , 
\end{equation*}
with a given jump probability 
\begin{displaymath}
  \Phi_r(\omega,\omega') = \lambda_r \, \phi_r(\omega,\omega').
\end{displaymath}
Here, $\phi_r$ satisfies the same normalization condition (\ref{eq:transition_normalization}) as the trail recruitment jump transition probability and $\lambda_r$ is the random velocity jump frequency.  With (\ref{eq:Qk2}), we find the expression of $Q_r$:
\begin{equation*}
  Q_r(F) = \lambda_r \, \left( \int \phi_r(\omega.\omega') F(x, \omega', t) \frac{d \omega'}{2 \pi} - F(x, \omega, t) \right) .
\end{equation*}
If $\phi_r = 1$, then $Q_r$ reduces to
\begin{equation*}
  Q_r(F) = \lambda_r \, \left( \frac{\rho(x,t)}{2 \pi} - F(x, \omega, t) \right) .
\end{equation*}

\medskip
\noindent
{\bf Summary of the kinetic model.} Below, we collect all equations of the kinetic model. We have written the model in the framework of  non-local interaction, preferential choice and nematic interaction. The restriction to simpler rules is easily deduced.
\begin{eqnarray}
  & & \hspace{-1cm} 
  \partial_t G(x, \omega, t) = \nu_d F(x, \omega, t) - \nu_e G(x, \omega, t), \label{eq:kin_phero}\\
  & & \hspace{-1cm} 
  \partial_t F + c \, \omega \cdot \nabla_x F = Q_r(F) + Q_p(F), \label{eq:kin_ant}\\
  & & \hspace{-1cm} 
  Q_p(F) (x,\omega, t) = \lambda_p \gamma (T_R(x,t)) \int_{\mathbb{S}^1} \phi_p(\omega,\omega') \big( h_p(\omega) \, F(x, \omega', t) \nonumber \\
  & & \hspace{7cm} 
  - h_p(\omega') \, F(x, \omega, t)\big) d \omega', \label{eq:kin_recruit}\\
  & & \hspace{-1cm} 
  Q_r(F) (x,\omega, t) = \lambda_r \int_{\mathbb{S}^1} \phi_r(\omega,\omega') \big(  F(x, \omega', t) -  F(x, \omega, t)\big) d \omega', \label{eq:kin_random}\\
  & & \hspace{-1cm} 
  h_p(\omega) = (g_R^{(sym)})^{[k]} (\omega), \quad g_R^{(sym)}(x,\omega,t) =  \frac{S_R(x,\omega,t) + S_R(x,-\omega,t)}{2T_R(x,t)}, \label{eq:kin_equi}\\
  & & \hspace{-1cm} 
  S_R(x, \omega, t) = \frac{1}{\pi R^2} \int_{|x-y|<R} G(y, \omega, t) d y, \quad T_R(x, t) = \int_{\mathbb{S}^1} S_R(x, \omega, t) \, d \omega. \label{eq:kin_sensing}
\end{eqnarray}
In the following section, we consider fluid limits of the present kinetic model.


\subsection{Macroscopic model}
\label{subsec:macroscopic}

{\bf Scaling.} In order to study the macroscopic limit of the kinetic model (\ref{eq:kin_phero})-(\ref{eq:kin_sensing}), we use the local interaction approximation $R = 0$, with non-nematic interaction and uniform transition probabilities $\phi_r = 1$, $\phi_p = 1$. In this case, the model simplifies into
\begin{eqnarray}
  & & \hspace{-1cm} 
  \partial_t G = \nu_d \, F - \nu_e \, G, \label{eq:kin_phero2}\\
  & & \hspace{-1cm} 
  \partial_t F + c \, \omega \cdot \nabla_x F = Q_r(F) + Q_p(F), \label{eq:kin_ant2}\\
  & & \hspace{-1cm} 
  Q_p(F) = \lambda_p \, \gamma (T) \, \left[ \rho \, h   - F \right], \label{eq:kin_recruit2}\\
  & & \hspace{-1cm} 
  Q_r(F) = \lambda_r \, \left( \frac{\rho}{2 \pi} - F \right) ,
  \label{eq:kin_random2}\\
  & & \hspace{-1cm} 
  h = g^{[k]}, \quad g = \frac{G}{T}, \quad T = \int_{\mathbb{S}^1} G \, d \omega, \quad  \rho = \int_{\mathbb{S}^1} F \,  d \omega, \label{eq:kin_rho2}
\end{eqnarray}
where the meaning of the power $[k]$ operation has been defined at (\ref{eq:(k)operation}). We now change to dimensionless variables. We let $t_0$, $x_0$, $\rho_0$, $T_0$, be respectively units of time, space, ant density and pheromone density and we introduce $x' = x/x_0$, $t' = t/t_0$, $\rho' = \rho/\rho_0$, $T' = T/T_0$, $F' = F/\rho_0$, $G' = G/T_0$ as new variables and unknowns. Specifically, $t_0$ is chosen to be the macroscopic time scale (e.g. the observation time scale). Similarly, $x_0$ is the macroscopic length scale (e.g. the size of the experimental arena). We impose $ x_0 = c t_0$, so that the time and space derivatives in (\ref{eq:kin_ant2}) are of the same orders of magnitude. This scaling allows us to observe the system at the convection scale where the convection speed of the ant density is finite.

We introduce the following dimensionless parameters: 
\begin{displaymath}
  \bar{\nu}_d  = \nu_d \, t_0, \quad \bar{\nu}_e  = \nu_e \, t_0 \frac{T_0}{\rho_0}, \quad \bar{\lambda}_p  = \lambda_p \, t_0, \quad \bar{\lambda}_r  = \lambda_r \, t_0.
\end{displaymath}
We make the assumption that the macroscopic time scale $t_0$ is very large compared to the microscopic time scales $\lambda_r^{-1}$ and $\lambda_p^{-1}$ which are both supposed to be of the same orders of magnitude. Indeed, during the time needed for patterns to develop, ants make a large number of jumps of either kind. Following this assumption, we introduce: 
\begin{displaymath}
  \varepsilon = \frac{1}{\bar{\lambda}_p} = \frac{1}{\lambda_p t_0}  \ll 1, \quad \sigma = \frac{\bar{\lambda}_r}{\bar{\lambda}_p}  = \frac{\lambda_r}{\lambda_p}  = O(1).
\end{displaymath}

Concerning the pheromone dynamics, we assume that $\bar{\nu}_d$ and $\bar{\nu}_e$ are of the same orders of magnitude, which amounts to supposing that pheromone deposition and evaporation balance each other. Indeed, if one of these two antagonist phenomena predominates, then, after some transient the pheromone density will become either too low or too large and we cannot expect any interesting patterns to emerge in this case. We introduce 
\begin{displaymath}
  \eta = \frac{1}{\bar{\nu}_d} = \frac{1}{\nu_d \, t_0}, \quad \kappa = \frac{\bar{\nu}_e}{\bar{\nu}_d} = \frac{\nu_e}{\nu_d} = O(1).
\end{displaymath}
In what follows, we will assume that $\eta = O(1)$ i.e.  that the pheromone dynamics occurs at the macroscopic time scale. 

After rescaling, system (\ref{eq:kin_phero2})-(\ref{eq:kin_rho2}) becomes (dropping the primes for the sake of clarity): 
\begin{eqnarray}
  & & \hspace{-1cm} 
  \eta \, \partial_t G^\varepsilon =  F^\varepsilon - \kappa \, G^\varepsilon , \label{eq:kin_phero3}\\
  & & \hspace{-1cm} 
  \varepsilon \, (\partial_t F^\varepsilon + \omega \cdot \nabla_x F^\varepsilon) = Q(F^\varepsilon), \label{eq:kin_ant3}
\end{eqnarray}
with the collision operator $Q = Q_r+Q_p$ given by 
\begin{eqnarray}
  & & \hspace{-1cm} 
  Q(F) = (Q_r+Q_p)(F) =  (\gamma(T) + \sigma) \, ( \mu \, \rho   - F ), \label{eq:kin_colli3}\\
  & & \hspace{-1cm} 
  \mu = \frac{\gamma(T) h + \frac{\sigma}{2 \pi}}{\gamma(T) + \sigma}, \quad
  h = g^{[k]}, \quad g = \frac{G}{T}, \label{eq:kin_sensing3} \\
  & & \hspace{-1cm} 
  T = \int_{\mathbb{S}^1} G \, d \omega, \quad  \rho = \int_{\mathbb{S}^1} F \,  d \omega. \label{eq:kin_rho3}
\end{eqnarray}

\medskip
\noindent
{\bf Macroscopic limit $\varepsilon \to 0$ of the kinetic model (\ref{eq:kin_phero3})-(\ref{eq:kin_rho3}).} Here, we suppose that $\eta = O(1)$ i.e. we assume that the pheromone dynamics occurs at the macroscopic scale. We show that the limit $\varepsilon \to 0$ of (\ref{eq:kin_phero3})-(\ref{eq:kin_rho3}) consists of the following system for the ant density $\rho(x,t)$, pheromone density $T(x,t)$ and pheromone distribution function $g(x,\omega,t)$: 
\begin{eqnarray}
  & & \hspace{-1cm} 
  \partial_t \rho + \nabla_x \cdot \left( \frac{\gamma(T)}{\gamma(T) + \sigma} \, j_h \right) = 0.,
  \label{eq:rho}
  \\
  & & \hspace{-1cm} 
  \eta \partial_t T = \rho - \kappa T,
  \label{eq:T} \\
  & & \hspace{-1cm} 
  \eta \partial_t g = \frac{\rho}{T} \left(\frac{\gamma(T) g^{[k]} + \frac{\sigma}{2 \pi}}{\gamma(T) + \sigma} - g\right),
  \label{eq:g_2} 
\end{eqnarray}
with $h = g^{[k]}$ and where $j_\varphi = \int_{{\mathbb S}^1} \varphi(\omega) \, \omega \, d\omega, $ denotes the flux of any function $\varphi(\omega)$. Eq. (\ref{eq:g_2}) is a closed equation for $g$. Once $g$ is determined and inserted into (\ref{eq:rho})  the evolution of the ant density $\rho$ can be computed. The ant distribution function $f$ is equal to $\mu$ at any time, with $\mu$ given by (\ref{eq:kin_sensing3}).

Indeed, in this limit, supposing that $F^\varepsilon \to F$, we get $Q(F) = 0$ from (\ref{eq:kin_ant3}). Therefore, from (\ref{eq:kin_colli3}), we obtain
\begin{equation} F = \rho \mu, \quad \quad \mbox{or} \quad \quad f = \mu. 
  \label{eq:equi_ants}
\end{equation}
The equation for $\rho(x,t)$ is obtained by integrating (\ref{eq:kin_ant3}) with respect to ${\omega}$ and using (\ref{eq:int_Qk=0}). We find:
\begin{displaymath}
  \partial_t \rho + \nabla_x \cdot j_F = 0.
\end{displaymath}
Remarking that $j_F = \rho j_\mu$ and that the flux of the isotropic distribution vanishes, we finally get from (\ref{eq:kin_sensing3}): 
\begin{equation} 
  j_\mu = \frac{\gamma(T)}{\gamma(T) + \sigma} \, j_h , 
  \label{eq:fluxf}
\end{equation}
and consequently, $\rho$ satisfies (\ref{eq:rho}). To compute the pheromone distribution function $g$, we integrate (\ref{eq:kin_phero3}) with respect to $\omega$ and get (\ref{eq:T}). Then, combining (\ref{eq:T}) with (\ref{eq:kin_phero3}), we deduce that 
\begin{equation} 
  \eta \partial_t g = \frac{\rho}{T} (f - g). 
  \label{eq:g}
\end{equation}
But, with (\ref{eq:equi_ants}) and (\ref{eq:kin_sensing3}), we deduce that $g$ satisfies (\ref{eq:g_2}).

Some comments are now in order. In the limit $\varepsilon \to 0$ the ant distribution function instantaneously relaxes to the distribution  $\mu$. This distribution reflects the antagonist effects of trail recruitment and random velocity jumps. Indeed, $\mu$ is the convex combination of the equilibrium distributions $h$ and $\frac{1}{2 \pi}$ of the two processes respectively. The weights, respectively equal to $\gamma(T)/(\gamma(T) + \sigma)$ and $\sigma/(\gamma(T) + \sigma)$ show that the influence of the trail recruitment process is more pronounced at large pheromone densities, since $\gamma$ increases with $T$. On the other hand, if the frequency of random jump $\sigma$ is increased, the trail recruitment process is comparatively less important. 

\medskip
\noindent
{\bf Case $k=1$: no preferential choice.} If the ants do not implement a preferential choice of the largest trails, i.e. if $k=1$, eq. (\ref{eq:g_2}) simplifies into 
\begin{eqnarray}
  & & \hspace{-1cm} 
  \eta \partial_t g = \frac{\rho}{T} \frac{\sigma}{\gamma(T) + \sigma} \, \left( \frac{1}{2 \pi} - g \right). 
  \label{eq:g_2_k=1}  
\end{eqnarray}
This is a classical relaxation equation of $g$ towards the isotropic distribution $\frac{1}{2\pi}$. As a consequence, in this case, there is no trail formation and the large time behavior of the system leads to a homogeneous steady state. This description can be complemented by looking at the pheromone flux $j_g$. Indeed, (\ref{eq:g_2_k=1}) leads to
\begin{displaymath}
  \eta \partial_t j_g = - \frac{\rho}{T} \frac{\sigma}{\gamma(T) + \sigma} \, j_g.
\end{displaymath}
As a consequence, the direction of the local pheromone flux never changes and its intensity decays to $0$ as $t \to \infty$. Additionally, eq. (\ref{eq:fluxf}) which in the case $k=1$ gives $ j_\mu = \frac{\gamma(T)}{\gamma(T) + \sigma} \, j_g$ shows that the ant flux is always proportional to and smaller than the pheromone flux. Therefore, it also converges to $0$ for large times. Note that this direction may not correspond to the maximum of the pheromone distribution $g$. Therefore, the ant flux may not be aligned with any particular trail, defined as such a maximum. 

The ant distribution $\mu$ is just the convex combination of the pheromone distribution $g$ and of the isotropic distribution. Therefore, the ant distribution is always smoother than the pheromone distribution. The random velocity jump process, even if very weak, seems to prevent a positive feedback between the ant and pheromone distributions which could lead to the formation of trails. 
Of course, these conclusions hold only when $\varepsilon \to 0$, i.e. if the equilibrium of the ant jump operator is instantaneously reached. The fact that the simulations do indeed show the formation of trails without any implementation of a preferential choice seems to indicate that the fast microscopic dynamics plays an important role in the formation of trails which the macroscopic model is unable to capture. We also note that if $\sigma = 0$, the pheromone distribution is constant in time. This is due to the fact that, in the absence of random velocity jumps, newly created pheromones are deposited according to a distribution which coincides exactly with the current pheromone distribution, resulting in an exact zero balance for this distribution. Therefore, even if $\sigma = 0$, no trails can develop.

\medskip
\noindent
{\bf Case $k>1$: existence of a preferential choice.} In this case, Eq. (\ref{eq:g_2}) is a non-local equation due to the operator $g \to g^{[k]}$. No analysis is available yet (to our knowledge) for such an equation (some preliminary results can be found in \cite{boissard_degond}). 
The large-time behavior of the system depends on the limit as $t \to \infty$ of eq. (\ref{eq:g_2}). We note that (\ref{eq:g_2}) may produce concentrations \cite{boissard_degond}. Indeed, the contribution of the largest trails is amplified and the ant flux becomes more strongly correlated to the direction of the largest trails. Therefore if the ants choose preferably the largest trail, the resulting concentration dynamics may counterbalance the  effect of the random velocity jumps and a positive feedback between the ants and the pheromones is more likely to occur. The study of this case is deferred to future work.

\medskip
\noindent
{\bf Conclusion on macroscopic models.} We have shown that macroscopic models are unable to develop trail formation without some mechanism allowing to amplify the variations of the pheromone distribution function. We have provided an example of such a mechanism, referred to as the preferential choice and which consists for the ants to choose the strong trails with higher probability than the weak ones. However, the need for such an amplification mechanism is not observed on the simulations of the microscopic model (see section \ref{sec:simulation}). This difference may indicate that the use of such macroscopic models is not fully justified for this dynamics. In particular, the chaos property (see e.g. \cite{cercignani_mathematical_1994}), which is the corner stone of the derivation of macroscopic models, may not be valid. Further rigorous mathematical studies are needed to make this point clearer.

\setcounter{equation}{0}
\section{Conclusion}
\label{sec:conclu}

In this article, we have introduced an Individual-Based Model of ant-trail formation. The ants are modeled as self-propelled particles which deposit directed pheromones (or pieces of trails) and interact with them through alignment interaction. We have introduced a trail detection technique which provides numerical evidence for the formation of trail patterns, and allowed us to quantify the effects of the biological parameters on the pattern formation. Finally, we have proposed both kinetic and fluid descriptions of this model and analyzed the capabilities of the fluid model to develop trail patterns. From the biological viewpoint, the model can be further improved. The ant and pheromone dynamics can be complexified for instance by adding extra pheromone diffusion, anisotropy or saturation in the pheromone detection mechanism, or by investigating the effect of a non-homogeneous medium. From the mathematical viewpoint, a rigorous derivation of the kinetic and fluid equations are still open problems.


\end{document}